\documentclass[aps,twocolumn]{revtex4}
\usepackage{graphicx}
\usepackage{color}
\newcommand{\Ket}[1]{|#1\rangle}

\begin{document}
\title{{\bf Three Level Atom Optics via the Tunneling Interaction}}
\author{K.~Eckert$^{a}$, M.~Lewenstein$^{a}$, R.~Corbal\'{a}n$^{b}$, G.~Birkl$^{c}$,
W.~Ertmer$^{c}$, and J.~Mompart$^{b}$}
\date{\today}
\address{$^{a}$ Institute of Theoretical Physics, University of Hannover, Appelstr.
2, D-30167 Hannover, Germany}
\address{$^{b}$ Departament de F\'{\i}sica, Universitat Aut\`{o}noma de Barcelona,
E-08193 Bellaterra, Spain}
\address{$^{c}$ Institute of Quantum Optics, University of Hannover, Welfengarten 1,
D-30167 Hannover, Germany}

\begin{abstract}
Three level atom optics (TLAO) is introduced as a simple, efficient and robust 
method to coherently manipulate and transport neutral atoms.
The tunneling interaction among three trapped states allows to realize the spatial analog
of the stimulated Raman adiabatic passage (STIRAP), coherent population trapping (CPT), 
and electromagnetically induced transparency (EIT) techniques. 
We investigate a particular implementation in optical microtrap arrays and show
that under 
realistic parameters the coherent manipulation and transfer of neutral atoms among dipole 
traps could be realized in the millisecond range. 
\end{abstract}


\maketitle

The coherent coupling between two orthogonal states of a quantum system gives rise
to oscillations of their probability amplitudes such as the Rabi
oscillations of a two-level atom interacting with a laser field. 
When three instead of two levels are considered, the interaction
gives rise to a much richer phenomenology.
A clear example is the
electric dipole interaction between a three-level atom and two laser modes,
where a large number of techniques have been proposed and reported, such as the 
stimulated Raman adiabatic passage (STIRAP) method used to produce a complete population
transfer between two internal quantum states of an atom or molecule \cite
{STIRAP}, the modification of the optical properties of a medium by
means of coherent population trapping (CPT) \cite{CPT}, and
electromagnetically induced transparency (EIT) \cite{EIT} phenomena. All
these {\it three-level optics} (TLO) techniques have been intensively studied
with applications ranging from quantum control of atoms and molecules \cite{STIRAP,control},
laser cooling \cite{lasercool}, and
slowing down light to a few meters per second \cite{slowlight} 
to non-linear optics with few photons \cite{fewphotons}.

In this letter we propose several novel techniques for the coherent
manipulation of atoms among trapped states coupled via tunneling. 
To illustrate the basic idea let us start by considering two well
separated dipole traps and one single atom in, say, the left trap. 
As soon as the two traps are approached and tunneling takes place, the
probability amplitude for the atom to be in the left 
(right)
trap oscillates in a Rabi-type fashion resembling the coupling of a two-level
atom to a coherent field. This tunneling induced oscillation between the two traps  
can be used to coherently transfer atoms between traps and, in fact, 
it allows for a simple realization of quantum computation \cite{SDQ}. 
However, this two-level technique is not very robust under variations 
of the system parameters and  requires precise control of distance and timing.  
We will introduce here a set of tools analogous to the TLO techniques 
to efficiently and coherently manipulate and move atoms among traps.
The basic elements will be three traps and a single atom, and
the atomic external degrees of freedom will be controlled through the variation of the
distance between each two traps. The proposed techniques do
not need an accurate experimental control of the system parameters
and they will be named 
{\it three level atom optics} (TLAO) techniques.

We will consider here arrays of optical microtraps
where the dipole force of a red detuned laser field is used to store neutral atoms
in each of the foci of a set of microlenses\cite{BirklOC}. We
will make use of two specific features of these arrays \cite{BirklPRL1}: 
the possibility of individual addressing each trap and detecting 
whether a trap is occupied, and the independent displacement of columns or rows 
of microtraps. 
We assume here that we are able to
initially store none or one atom per trap at will, as has been reported in
single dipole traps \cite{singledt} and in optical lattices \cite
{singleMott}. Although we require only three traps the use
of columns of traps has the advantage of doing several
experiments in parallel.

\begin{figure}
\begin{center}
\includegraphics[width=7cm,clip]{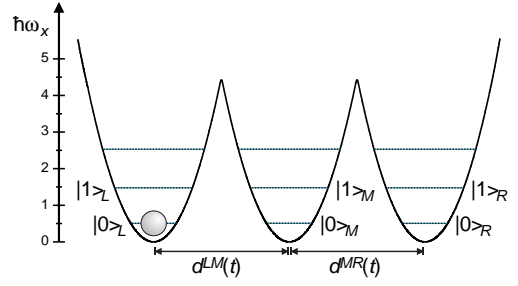}
\end{center}
\caption[]{ 
Three trap potential; $d^{LM}$ ($d^{MR}$) is the separation between
left and middle (middle and right) traps. In the limit of a large separation
$\left| n \right\rangle _{L},\;\left| n \right\rangle _{M},\;\left| n \right\rangle _{R}$ 
are the $n$th vibrational energy eigenstates of the corresponding single trap potentials.
}
\end{figure}

The three in-line dipole traps are modeled as three piece-wise harmonic
potentials 
of frequency $\omega_x$. and the neutral atom is assumed to be in the ground vibrational state
of the left trap initially, while the other two traps are empty (Fig.~1). 
For simplicity the temporal evolution of the distance between each two traps 
has been modeled with a cosine function truncated at the minimum separation.
Then, the approach and eventual separation of left 
and middle (middle and right) traps takes a time $t_{r}^{LM}$ ($t_{r}^{MR}$),
while $t_{i}^{LM}$ ($t_{i}^{MR}$) is the time the traps remain at the minimum distance.
The (unperturbed) three-level system is composed of the vibrational 
ground states of all three traps, i.e., $\left| 0\right\rangle _{L}$, $\left| 0\right\rangle _{M}$%
, and $\left| 0\right\rangle _{R}$, and the strength of the interaction
between each two vibrational ground states is given (in the absence of the third trap) 
by the following tunneling ''Rabi'' frequency \cite{Rabi}: 
\begin{equation}
\Omega (d) = \frac{-1+e^{(\alpha d)^{2}}\left[ 1+\alpha d\left( 1-%
%TCIMACRO{\func{erf}}%
%BeginExpansion
\mathop{\rm erf}%
%EndExpansion
(\alpha d)\right) \right] }{\sqrt{\pi }\left( e^{2(\alpha d)^{2}}-1\right) /2 \alpha d}
\end{equation}
where $\alpha d$ is the trap separation, and $\alpha^{-1} \equiv \sqrt{\hbar/m \omega_x}$ 
with $m$ denoting the mass of the neutral atom. ${\rm erf}(.)$ is the error function. 
The temporal shaping of $\Omega $ is realized by controlling 
the time dependence of $d(t)$.
While Eq.~(1) is useful to explore the analogies between TLO and TLAO,
an exact treatment accounting also for couplings to excited vibrational states requires the 
integration of the Schr\"odinger equation. In what follows we will numerically 
integrate the 1D, and, eventually, the 2D Schr\"odinger equation to 
simulate the dynamics of the neutral atom in the three-trap potential. 

\begin{figure}
\begin{center}
\includegraphics[width=9cm]{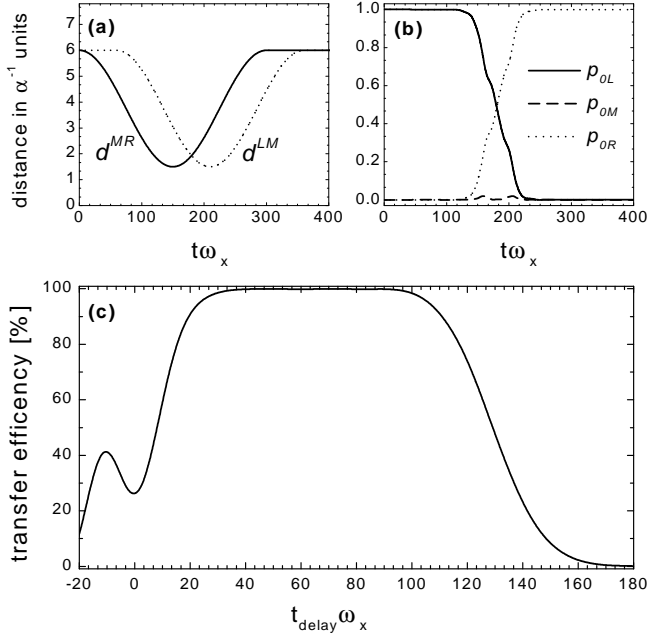}
\end{center}
\caption[]{
(a) Approaching sequence for a STIRAP-like process, 
and (b) the corresponding ground state populations; 
$d^{LM}_{max}\alpha=d^{MR}_{max}\alpha=6$, 
$d^{LM}_{min}\alpha=d^{MR}_{min}\alpha=1.5$,
$t^{LM}_{r}\omega_x=t^{MR}_{r}\omega_x=150$,
$t^{LM}_{i}\omega_x=t^{MR}_{i}\omega_x=0$,
and $t_{delay}\omega_x=60$.
(c) Transfer efficiency from  $\left| 0\right\rangle _{L}$ to $\left| 0\right\rangle _{R}$ 
as a function of the time delay between the two approaching processes.
}
\end{figure}

A robust method to coherently move atoms among traps
consists in extending the STIRAP technique \cite{STIRAP} to atom optics by using
the tunneling interaction. The basic idea is to use the fact that one of the
three eigenstates of the three level system involves only the ground states
of the two extreme traps: 
\begin{equation}
\left| D ( \Theta )\right\rangle \equiv
\cos \Theta \left| 0\right\rangle _{L}-\sin \Theta
\left| 0\right\rangle _{R},
\end{equation}
where the mixing angle $\Theta $ is defined as $\tan \Theta \equiv \Omega
^{LM}/\Omega ^{MR}$ with $\Omega ^{LM}$ ($\Omega ^{MR}$) denoting 
the tunneling "Rabi" frequency between
left and middle (middle and right) traps.
Following Eq.~(2) it is possible to transfer the atom from $%
\left| 0\right\rangle _{L}$ to $\left| 0\right\rangle _{R}$ by adiabatically
varying the mixing angle from $0^{\circ }$ to $90^{\circ }$, which
means to approach and separate first the right trap to the middle one and,
with an appropriate delay, the left trap to the middle one (Fig.~2(a)). This
counterintuitive sequence moves the atom directly from $\left|
0\right\rangle _{L}$ to $\left| 0\right\rangle _{R}$ with an almost
negligible probability amplitude to be in the middle trap ground state (Fig.~2(b)). 
The STIRAP signature is shown in Fig.~2(c), where the transfer efficiency from $%
\left| 0\right\rangle _{L}$ to $\left| 0\right\rangle _{R}$ is shown as a
function of the time delay between the two approaching processes. The plateau
near the optimal delay indicates the robustness
of the transfer process which also is very robust under variations of the tunneling parameters, i.e., of the 
maximum and minimum trap separation $d_{max}$ and $d_{min}$, and of $t_{r}$ and $t_{i}$
for each of the processes, provided that adiabaticity is maintained. 

\begin{figure}
\begin{center}
\includegraphics[width=9cm]{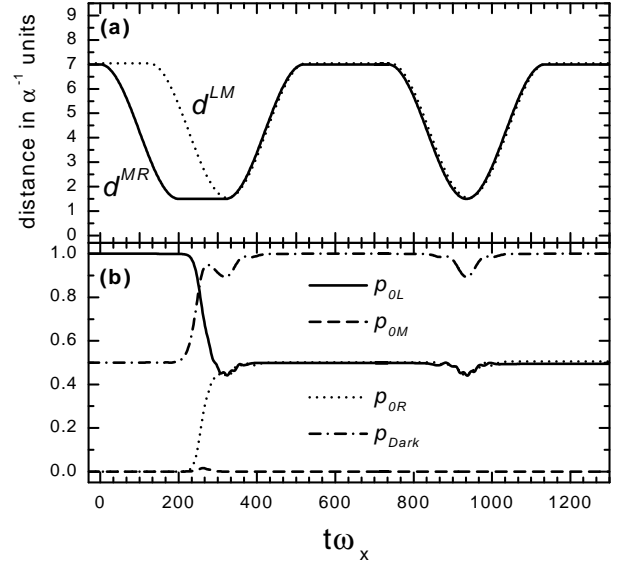}
\end{center}
\caption[]{
(a) Approaching sequence for a CPT-like process; (b) Ground state and dark state populations, 
$p_{Dark}=$ $\left| \left\langle D (\Theta = \pi /2) | \psi (t) \right\rangle \right|^2$;
$d^{LM}_{max}\alpha=$ $d^{MR}_{max}\alpha=7$, 
$d^{LM}_{min}\alpha=$ $d^{MR}_{min}\alpha=1.5$,
$t^{LM}_{r}\omega_x=$ $t^{MR}_{r}\omega_x=200$,
$t^{LM}_{i}\omega_x=0$, $t^{MR}_{i}\omega_x=t_{delay}\omega_x=120$
for the processes occurring before $t \omega_x = 600$.
From $t \omega_x = 600$ on the parameters are the same 
as in the previous case except for
$t^{MR}_{i}\omega_x=t_{delay}\omega_x=0$.
}
\end{figure}

Additionally, the approaching sequence can be modified to create spatial superposition
states with maximum atomic coherence, i.e., with $\left| c_{0L}
c_{0R}^* \right| =1/2$, $c_{0L}$ ($c_{0R}$) being the probability amplitude 
to be in state $\left| 0\right\rangle _{L}$ ($\left| 0\right\rangle _{R}$).
The basic idea is to adiabatically following state (2) from $\Theta = 0^{\circ }$ 
up to $\Theta = 45^{\circ }$ by an appropriate delay in the approaching process and the subsequent
simultaneous separation of the outermost traps (see Figs.~3(a) and 3(b) up to 
$t\omega_x = 600$). The resulting state is the spatial equivalent to the well known dark state 
arising in the CPT technique \cite{CPT}. To prove that this state
is dark, i.e., that it can be decoupled from the tunneling interaction, we approach 
and separate simultaneously the two extreme traps to the middle one 
(see Figs.~3(a) and 3(b) from $t\omega_x = 600$ up to the end). Clearly, the atom remains 
in the dark state in spite of the tunneling interaction. 
Superposition dark states are very sensitive to dephasing \cite{CPT}
which means that they could be used 
in dipole trap systems to measure experimental imperfections 
such as uncorrelated shaking in the traps position and/or intensity fluctuations 
of the trapping lasers. 
Also, this robust coherent splitting of the atomic wave function into two half pieces together with their 
individual manipulation anticipates applications in atomic interferometry.

\begin{figure}
\begin{center}
\includegraphics[width=9cm]{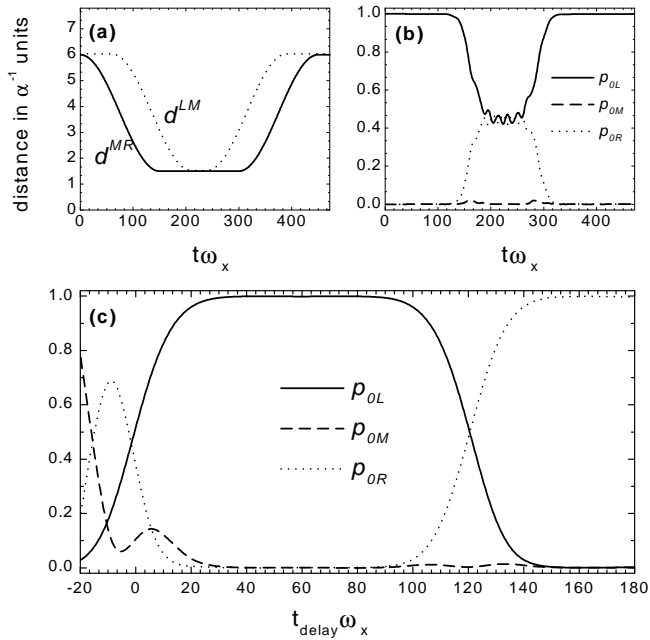}
\end{center}
\caption[]{
(a) Approaching sequence for an EIT-like process,
and (b) corresponding ground state populations;
$d^{LM}_{max}\alpha=d^{MR}_{max}\alpha=6$, 
$d^{LM}_{min}\alpha=d^{MR}_{min}\alpha=1.5$,
$t^{LM}_{r}\omega_x=$ $t^{MR}_{r}\omega_x=150$,
$t^{LM}_{i}\omega_x=30$, 
$t^{MR}_{i}\omega_x=150$,
and $t_{delay}\omega_x=60$.
(c) Ground state populations as a function of 
the time delay.
}
\end{figure}

Finally, it is also possible to extend the EIT technique \cite{EIT} to the atom optics case. 
The basic idea of EIT is to convert a medium that is opaque to
a field resonant with a certain internal transition into being transparent
by applying an intense driving field to an adjacent transition. 
In the three-trap system we will inhibit the transition 
from $\left| 0\right\rangle _{L}$ to $\left| 0\right\rangle _{M}$
by driving the transition $\left| 0\right\rangle _{M}$ $\leftrightarrow$
$\left| 0\right\rangle _{R}$ via the tunneling interaction. Figs.~4(a) and 4(b) show the inhibition
of the $\left| 0\right\rangle _{L}$ to $\left| 0\right\rangle _{M}$ transition
in spite of the proximity between left and middle traps. 
As in the STIRAP case, the plateau near the optimum delay
evidences the robustness of the transition cancellation (Fig.~4(c)). 
This atom optics EIT technique could create conditional phase shifts 
with no change in the state of the system.

\begin{figure}
\begin{center}
\includegraphics[width=8cm]{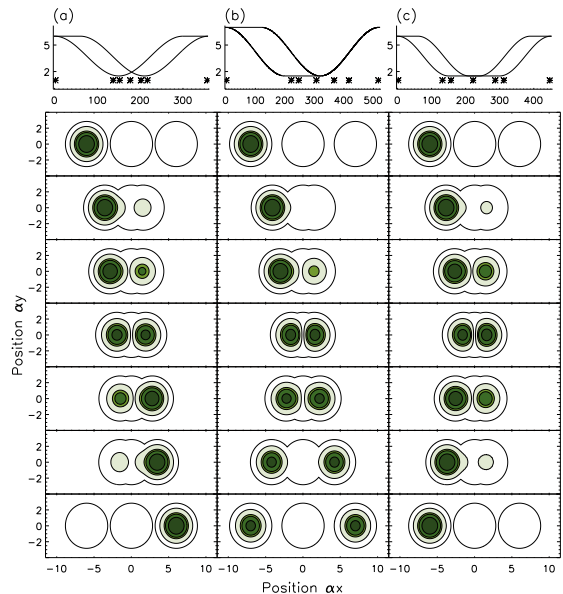}
\end{center}
\caption[]{2D probability distributions for the ground state spatial analogon of 
(a) STIRAP, (b) CPT, and (c) EIT. Parameters as in Figs.~2, 3 and 4, respectively, with $\omega_y=\omega_x$. 
Asterisks in the top plots indicate the time of the snapshots.}
\end{figure}

Fig.~5 shows the results of a 2D numerical integration of the Schr\"odinger equation 
that summarizes the previously discussed ground state TLAO techniques.
It is worth to note that these techniques can be also applied to excited states,
which relaxes the cooling requirements for the experimental setup.
In Fig.~6 two examples for the three level system consisting of the first excited
vibrational states of each trap are considered: 
(a) the transfer of an atom from $\Ket{1}_L$ to $\Ket{1}_R$
via the STIRAP technique, and (b) the 50\% coherent splitting of the atomic wavefunction
between the left and the middle trap (see the plateau around $t \omega_x = 180$). 
This effect, which is different from CPT and requires a combination of adiabatic
and diabatic processes \cite{fewell}, is possible through a complicated variation of the
dressed level structure of the first excited states when approaching the traps.

There are two important
practical points for the implementation of the TLAO techniques in optical microtrap arrays:
(i) the trapping frequencies must be the same for all microtraps; and  
(ii) the approaching process has to be adiabatic. 
The use of a single laser that illuminates simultaneously all microlenses
assures the identity of all microtraps even under intensity fluctuations of the laser. 
In particular, typical trapping frequencies for microtrap arrays of $^{87}$Rb atoms are 
10$^{5}$-10$^{6}$ s$^{-1}$ in the transverse directions and 10$^{4}$-10$^{5}$ s%
$^{-1}$ along the laser beam direction \cite{BirklOC,BirklPRL1} which means that the traps
can be adiabatically approached in the millisecond range or even faster
by using optimization techniques \cite{SDQ,optimization}. 
In addition, the spatial analogous of the CPT technique requires a precise control of the ratio 
between the two relevant "Rabi" frequencies. Fortunately, even in the presence of shaking in 
the microtraps position the ratio between the "Rabi" frequencies can be accurately controlled 
since mechanical vibrations give rise to a correlated shaking.

\begin{figure}
\begin{center}
\includegraphics[width=9cm]{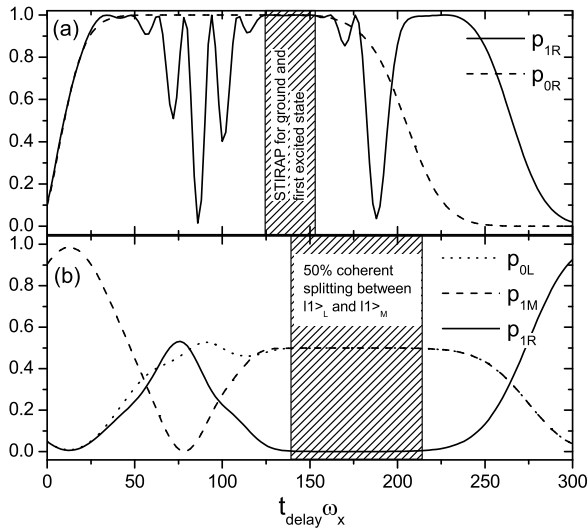}
\end{center}
\caption[]{Three level atom optics for the first excited vibrational states:
(a) transfer efficiency for STIRAP (dashed line: efficiency for the ground state STIRAP for the same parameters);
and (b) population for the coherent splitting between $\Ket{1}_L$ and $\Ket{1}_M$; the parameters are 
(a) $t^{LM}_{r}\omega_x=t^{MR}_{r}\omega_x=300$,
$t^{LM}_{i}\omega_x=t^{MR}_{i}\omega_x=0$,
(b) $t^{LM}_{r}\omega_x=550, t^{MR}_{r}\omega_x=400$,
$t^{LM}_{i}\omega_x=75, t^{MR}_{i}\omega_x=400$. In both cases
$d^{LM}_{max}\alpha=d^{MR}_{max}\alpha=9$ and 
$d^{LM}_{min}\alpha=d^{MR}_{min}\alpha=1.5$.
}
\end{figure}

Throughout the paper we have assumed to be able to cool down the atom 
to the lower vibrational states of the traps. 
In fact, sideband cooling to a temperature below  $1 \mu {\rm K}$ 
with a ground state population of $98.4$\% has been reported in optical lattices \cite{ground}
with parameters very similar to the ones considered here. In this case,  
heating rates below $1 \mu {\rm K} /{\rm s}$ have been estimated \cite{heating}.
In the presence of decoherence from heating, shaking and spontaneous scattering, 
fidelities above 98\% \cite{SDQ} can be expected for the ground-state TLAO techniques discussed here. 
However, it is worth to note again that all these techniques can be also applied to excited states.  
Note that the real trapping potentials differ from simple harmonic ones,
but the three level atom optic techniques discussed here do not rely on the 
particular shape of the trapping potentials, provided the adiabaticity 
is maintained during the whole process of approaching and separating the traps.

Summarizing, we have introduced a set of robust and efficient techniques to coherently
manipulate and transport neutral atoms based on three-level atom optics.
These atom optics techniques correspond to the natural extension of the largely investigated 
STIRAP, CPT and EIT techniques used in quantum optics
\cite{STIRAP,CPT,EIT} with the interaction mediated via tunneling 
and controlled by the shaping of the process of varying the separation between the traps.
The fact that three-level atom optics makes use of the tunneling interaction 
means some important differences with respect to the quantum optics case,  
such as the time scale of the processes being in the millisecond range, 
the absence of electric dipole rules, or the possible use of excited states.
Applications to atomic interferometry and precision measurement
have been briefly discussed and some practical considerations for the implementation
in dipole trap arrays have been addressed. These three-level atom
optics techniques are widely applicable also in other atom optics 
systems such as magnetic microtraps, optical lattices, 
and dipole and magnetic waveguides.

We would like to thank U.~Poulsen for useful discussions
and acknowledge financial support from MCyT (Spain) and FEDER (EC) under 
project BFM2002-04369, and from the DGR (Catalunya) under project 2001SGR-187, as well
as from CESCA-CEPBA, the DFG (SPP Quantum Information Processing and SFB 407),
and from ACQP (EC). 

%Research partially supported by the Improving the Human Potential Programme,
%Acces to Research Infraestructures, under contract HPRI-1999-CT-00071 
%"Access to CESCA and CEPBA Large Scale Facilities" 
%established between the European Community and CESCA-CEPBA.

\end{document}